\UseRawInputEncoding
\documentclass[pdflatex,sn-mathphys-num]{sn-jnl}

\usepackage{graphicx}%
\usepackage{multirow}%
\usepackage{amsmath,amssymb,amsfonts}%
\usepackage{amsthm}%
\usepackage{mathrsfs}%
\usepackage[title]{appendix}%
\usepackage{xcolor}%
\usepackage{textcomp}%
\usepackage{manyfoot}%
\usepackage{booktabs}%
\usepackage{algorithm}%
\usepackage{algorithmicx}%
\usepackage{algpseudocode}%
\usepackage{listings}%

\begin{document}

\title[Article Title]{Incommensurate Charge Super-modulation and Hidden Dipole Order in Layered Kitaev Material $\alpha$-RuCl$_3$}


\author*[1]{Xiaohu Zheng}\email{zhengxh@baqis.ac.cn}

\author*[2,3]{Zheng-Xin Liu}\email{liuzxphys@ruc.edu.cn}

\author[4]{Cuiwei Zhang}
\author[1]{Huaxue Zhou}
\author[1]{Chongli Yang}
\author[4]{Youguo Shi}
\author[1]{Katsumi Tanigaki}
\author*[5,6,7]{Rui-Rui Du}\email{rrd@pku.edu.cn}
\affil[1]{Beijing Academy of Quantum Information Sciences, Beijing 100193, China}

\affil[2]{Department of Physics and Beijing Key Laboratory of Opto-electronic Functional Materials and Micro-nano Devices, Renmin University of China, Beijing, 100872, China}

\affil[3]{Key Laboratory of Quantum State Construction and Manipulation (Ministry of Education), Renmin University of China, Beijing, 100872, China}

\affil[4]{Institute of Physics, Chinese Academy of Sciences, Beijing 100190, China}

\affil[5]{International Center for Quantum Materials, School of Physics, Peking University, Beijing 100871, China}

\affil[6]{CAS Center for Excellence in Topological Quantum Computation, University of Chinese Academy of Sciences, Beijing 100190, China}

\affil[7]{Hefei National Laboratory, Hefei 230088, China}


\abstract{The magnetism of Kitaev materials has been widely studied, but their charge properties and the coupling to other degrees of freedom are less known. Here we investigate the charge states of $\alpha$-RuCl$_3$, a promising Kitaev quantum spin liquid candidate, in proximity to graphite. We discover that few-layered $\alpha$-RuCl$_3$ experiences a clear modulation of charge states, where a Mott-insulator to weak charge-transfer-insulator transition in the 2D limit occurs by means of heterointerfacial polarization. More notably, distinct signals of incommensurate charge and lattice super-modulations, regarded as an unconventional charge order, accompanied in the insulator. Our theoretical calculations have reproduced the incommensurate charge order by taking into account the antiferroelectricity of $\alpha$-RuCl$_3$ that is driven by dipole order in the internal electric fields. The findings imply that there is strong coupling between the charge, spin, and lattice degrees of freedom in layered $\alpha$-RuCl$_3$ in the heterostructure, which offers an opportunity to electrically access and tune its magnetic interactions inside the Kitaev compounds.}




\maketitle

\section*{Introduction}

\begin{figure}[htp]
 	\centering
 	\includegraphics[width=1\columnwidth]{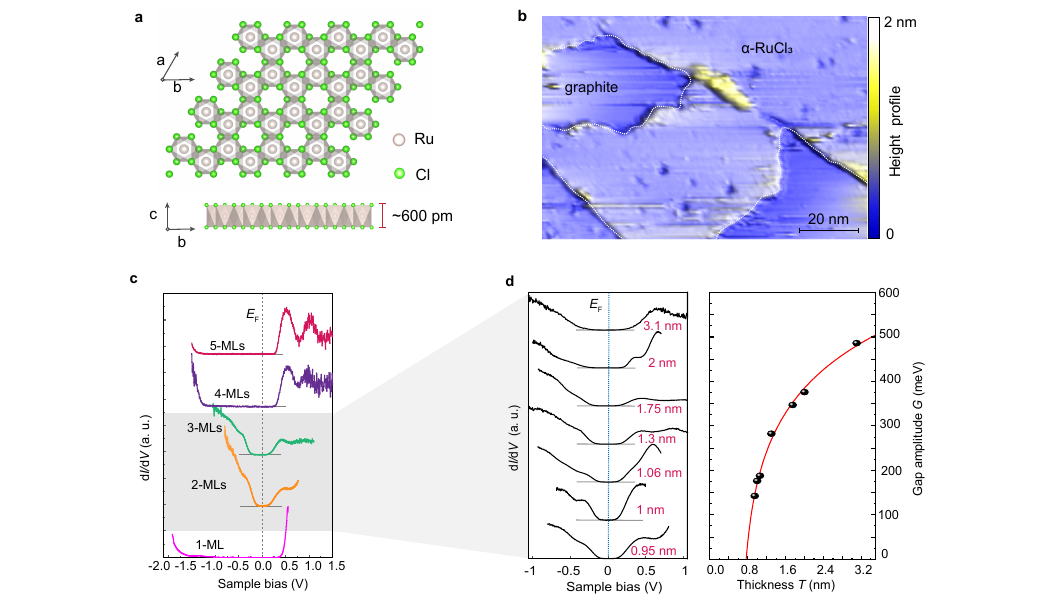}\\
 	\linespread{1}
 	\caption[Figure 1]{\textbf{Unconventional Mott transition in $\alpha$-RuCl$_3$ in proximity to graphite.} \textbf{a} Crystal structure of $\alpha$-RuCl$_3$ in the $ab$ and $bc$ planes; \textbf{b} STM morphology of $\alpha$-RuCl$_3$ flakes transferred on a graphite substrate; \textbf{c} Selected d$I$/d$V$ spectra taken on $\alpha$-RuCl$_3$ flakes with thicknesses ranging from 1-ML to 5-MLs showing two types of line-shapes, one with a large gap and the other with a reduced gap ($V_{\rm bias}$: 0.5-1 V, $I_{\rm set}$: 0.5-1 nA); \textbf{d} Left: averaged d$I$/d$V$ spectra taken on a series of $\alpha$-RuCl$_3$ thin flakes with different thicknesses. Right: Plots of the relationship between the gap-amplitude ($G$) and the appropriate thickness ($T$) expected from the logarithmic function $G=\alpha-\beta {\ln}(T+\gamma)$ (where $\alpha$, $\beta$, and $\gamma$ are constants).}
 	\label{fig:figure-1}
 \end{figure}

Quantum spin liquids (QSLs) are exotic Mott insulating phases without long-range magnetic order, but having long-ranged entanglements and fractionalized excitations \cite{anderson_resonating_1973,fazekas_ground_1974,balents_spin_2010, broholm_quantum_2020,zhou_quantum_2017}. The Kitaev honeycomb model, which hosts accurate ground state and Majorana excitations \cite{kitaev_anyons_2006, takagi_concept_2019, motome_hunting_2019}, attracted lots of interest in realization of QSLs. As a 
Kitaev QSL candidate \cite{jackeli_mott_2009,rau_generic_2014}, the Mott insulator (MI) $\alpha$-RuCl$_3$ (as shown in Fig. \ref{fig:figure-1}a) has been extensively investigated in the past decade \cite{plumb_ensuremathalphaensuremath-mathrmrucl_3_2014, kubota_successive_2015, kim_kitaev_2015, johnson_monoclinic_2015}. It has been shown that $\alpha$-RuCl$_3$ is proximate to the Kitaev paramagnet between 7 and 120 K \cite{do_majorana_2017,sandilands_scattering_2015,sears_phase_2017}, where a continuum was observed in spectrum of inelastic neutron scattering (INS) \cite{do_majorana_2017, banerjee_neutron_2017, banerjee_proximate_2016,banerjee_excitations_2018,ran_spin-wave_2017}, Raman scattering \cite{sandilands_scattering_2015,sandilands_spin-orbit_2016, wulferding_magnon_2020, lee_multiple_2021, yu_ultralow-temperature_2018, lin_anisotropic_2020} and other spectral techniques \cite{zheng_gapless_2017, sears_ferromagnetic_2020}. Furthermore, under external magnetic fields thermal transport measurements have detected half-quantization of the thermal Hall conductance \cite{yokoi_half-integer_2021, bruin_robustness_2022,kasahara_majorana_2018,imamura_majorana-fermion_2024} as well as quantum oscillation in the longitudinal thermal conductance \cite{czajka_oscillations_2021,villadiego_pseudoscalar_2021} despite the anomalous phonon contribution \cite{NC_phononAnomaly21,PRX_phononHall22}. These observations indicate that $\alpha$-RuCl$_3$ is an ideal platform to study fractionalized spin excitations.\par

Aside from the magnetism, the charge properties of Kitaev materials has attracted increasing interest \cite{PRB_Sunliling18,npjCM_Angel23,Nanomat_Kim24, mi_observation_2022}. Nevertheless, a significant concern is that these materials, like $\alpha$-RuCl$_3$, are electrically inert MIs, which prevent any low-energy charge excitations. Forming a heterostructure with metallic system has been demonstrated to be an excellent strategy for injecting active charges and triggering exotic phenomena. For instance, it was proposed that electrons tunneled from substrate can locally probe and control the magnetic fractional excitations in Kitaev thin film \cite{udagawa_scanning_2021,bauer_scanning_2023,pereira_electrical_2020, konig_tunneling_2020,feldmeier_local_2020}. Investigations of charge transfer and magnetic phase transition at the interface between $\alpha$-RuCl$_3$ and graphene \cite{NanoLetter_Kern18,biswas_electronic_2019,zhou_evidence_2019,rizzo_charge-transfer_2020,rizzo_nanometer-scale_2022,rossi_direct_2023} or MnPc \cite{PRB_Koitzsch22} were reported. Potentially, the coupling between the charge, spin, and lattice degrees of freedom are helpful to tune the magnetic interactions and to manipulate the neutral fractional excitations, as proposed in a series of theoretical papers \cite{aasen_electrical_2020,klocke_time-domain_2021,dhochak_magnetic_2010, vojta_kondo_2016,wang_emergent_2021,PRB_MacDonald23}. Therefore, it is an important issue how the Mott nature can be modified in Kitaev thin film on metallic surface.\par

Our early works have revealed an unconventional Mott transition in few-layer $\alpha$-RuCl$_3$ \cite{zheng_tunneling_2023,zheng_insulator--metal_2024}. Here, we report our new scanning tunneling microscopy/spectroscopy (STM/STS) experimental observations of incommensurate super-modulation of both charge density and lattice morphology on $\alpha$-RuCl$_3$ during its transition from MI to charge transfer insulator (CTI). Our results suggest that the CTI of few-layer $\alpha$-RuCl$_3$, with a significantly reduced charge gap, enables low-energy charge excitations which are strongly coupled to the spin and lattice degrees of freedom. Specifically, the anti-ferroelectric dipole order in few-layered $\alpha$-RuCl$_3$ has induced the unconventional charge order under intrinsic fluctuation here. Our work not only reveals the charge properties of the Kitaev material $\alpha$-RuCl$_3$, but also shed light on the understanding of charge-spin coupling in general quantum magnets.\par

\section*{Results}
We perform the STM/STS study of few layers $\alpha$-RuCl$_3$ that have been transferred onto the surface of graphite.  
A selected STM topography recorded at 77 K is shown in Fig.\ref{fig:figure-1}b. The spectra were systematically acquired on a series of thin flakes ranging from 1 to 5-MLs, as well as thicker flakes (Supplementary Fig. 1), where the thicknesses were measured through the STM cross-sectional profiles. Two types of representative d$I$/d$V$ spectra, both of which exhibit a full charge-gap, are shown in Fig. \ref{fig:figure-1}c. The first type, which belongs to the flakes of 4-MLs and above, bears resemblance to that of bulk (Supplementary Fig. 1 and references \cite{sinn_electronic_2016,nevola_timescales_2021}), i.e., the gap values remain close to 2 eV, revealing a strong insulating state. The second type is identified in few-layers that are situated in close proximity to graphite and exhibits a dramatically reduced charge-gap, which will be later referred to as the reduced-gap (RG). During the transition between the two gapped phases as decreasing the thickness, the d$I$/d$V$ spectrum of MI experiences a dramatic transfer of spectral weight from Hubbard bands (HBs) to the sides adjacent to the Fermi level (FL) (Supplementary Fig. 2). d$I$/d$V$ spectra pertaining to the RG phase were collected on a few dozens of thin flakes (see left in Fig.\ref{fig:figure-1}d), and we found a logarithmic decrease in the gap as the surface approached the substrate, as depicted in right panel of Fig.\ref{fig:figure-1}d. \par

\begin{figure}[htp]
 	\centering
 	\includegraphics[width=1\columnwidth]{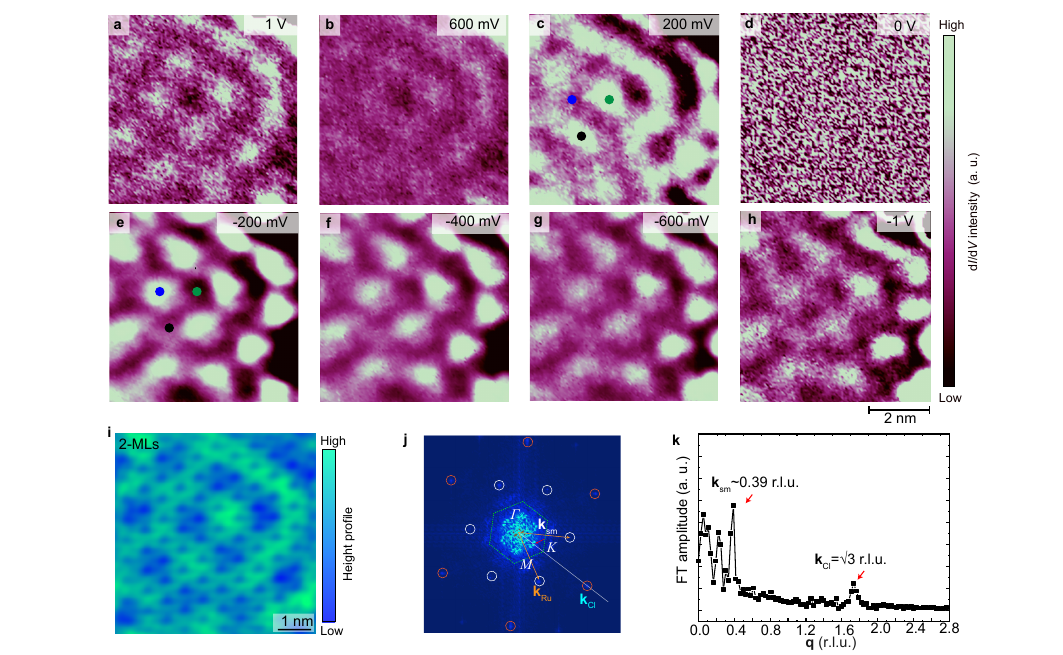}\\
 	\linespread{1}
 	\caption[Figure 2]{\textbf{Incommensurate charge order in $\alpha$-RuCl$_3$.} \textbf{a} to \textbf{h} show sample bias-resolved d$I$/d$V$ maps measured at $V_{\rm bias}$=500 mV, $I_{\rm set}$=1 nA on a 2-ML $\alpha$-RuCl$_3$. Color dots in (c) and (e) indicate the reversal patterns with weaker contrast at ±200 mV; \textbf{i} Atomic-resolved STM image recorded on the same place in (a-h) shows that the super-modulation is also observable on the lattice ($V_{\rm bias}$=500 mV, $I_{\rm set}$=1 nA); \textbf{j} FFT of the STM image shows the Bragg peaks of the $\alpha$-RuCl$_3$ lattice (white and red circles) and peaks of the static super-modulation inside the 1st BZ (green dashed hexagon). $\mathbf k$$_{\rm Ru}$ and  $\mathbf k$$_{\rm Cl}$ denote the wave vectors of ruthenium (Ru) and chlorine (Cl) lattice, respectively. $\mathbf k$$_{\rm sm}$ is the wave vector of the super-modulation; \textbf{k} The line cut along the $\Gamma$-K direction shows that $\mathbf k$$_{\rm sm}$ is approximately 0.39 r.l.u. (r.l.u. denotes reciprocal lattice unit of $\alpha$-RuCl$_3$ 2$\pi$/$a$, $a$ = 6 \AA).}
 	\label{fig:figure-2}
 \end{figure}  
 
 In order to further understand the electronic states of RG phase in $\alpha$-RuCl$_3$, we first focus on 2-MLs. The energy-dependent d$I$/d$V$ maps (Figs.\ref{fig:figure-2}a to \ref{fig:figure-2}h) were performed on an area exhibiting atomic-resolved STM morphology on $\alpha$-RuCl$_3$ surface, as shown in Fig.\ref{fig:figure-2}i. The most important feature is the super-modulation of the surface local density of state (LDOS), which resembles  a charge density wave order. The modulation is also observable on the lattice (see Fig.\ref{fig:figure-2}i). The findings presented in Supplementary Fig. 3 reveal that the super-modulation primarily occurs in the occupied states (negative sample biases). As depicted in Figs.\ref{fig:figure-2}e to \ref{fig:figure-2}g, it is most pronounced within the sample biases of -200 mV and -800 mV, and diminishes as the bias deviates from this range (refer to Supplementary Fig. 3c for further details). The shaded circles in Figs.\ref{fig:figure-2}c and e demonstrate a contrast reversal occurring at the band edges. Furthermore, it should be noted that while the amplitude of the super-modulation exhibits a strong correlation with the biases, the patterns remain relatively static with respect to the energy. \par  
 To further analyze the super-modulation, a fast Fourier transform (FFT) was performed. The Bragg peaks corresponding to both ruthenium (Ru) and chlorine (Cl) atoms are highlighted by the white and red circles in Fig.\ref{fig:figure-2}j. The set of peaks observed along $\Gamma$-$K$ direction within the first Brillouin zone (BZ) is a clear evidence of the static super-modulation present on the morphology. Upon a line cut along the white dashed line in Fig.\ref{fig:figure-2}j, it was determined that the wave vector $\mathbf k$$_{\rm sm}$ has a value of approximately 0.39 r.l.u (r.l.u. denotes reciprocal lattice unit of $\alpha$-RuCl$_3$ 2$\pi$/$a$, $a$ = 6 \AA) (Fig.\ref{fig:figure-2}k), indicating that the super-modulation is incommensurate.\par

Our systematic investigation confirms that such surface modulations are commonly occurring in 2-MLs. The wave vector $\mathbf k$$_{\rm sm}$ remains static and does not depend on the orientations between $\alpha$-RuCl$_3$ and graphite in different samples. In a second 2-MLs sample, the bias-dependent STM images exhibit a conventional $\alpha$-RuCl$_3$ lattice with moderate morphological modulation under positive biases (Supplementary Fig. 4a), similar to that shown in Fig.\ref{fig:figure-2}i. In contrast, the super-modulation arises and prevails throughout the surface under negative biases, which becomes particularly prominent around a bias of -600 mV, as shown in Supplementary Fig. 4a. Furthermore, we find the modulation on surface LDOS and around defects evolve almost synchronously with the morphological super-modulation under variation of biases, as shown in Supplementary Fig. 4b. The evolution is also revealed in the bias-resolved Fourier components of the d$I$/d$V$ data (Supplementary Fig. 4c), where the FFT patterns of the super-modulation become more prominent and evolve from point-like pattern to nearly ring-shape around the $\Gamma$ point as bias shifts towards the negative end (Supplementary Fig. 4c).\par
 
 \begin{figure}[htp]
 	\centering
 	\includegraphics[width=1\columnwidth]{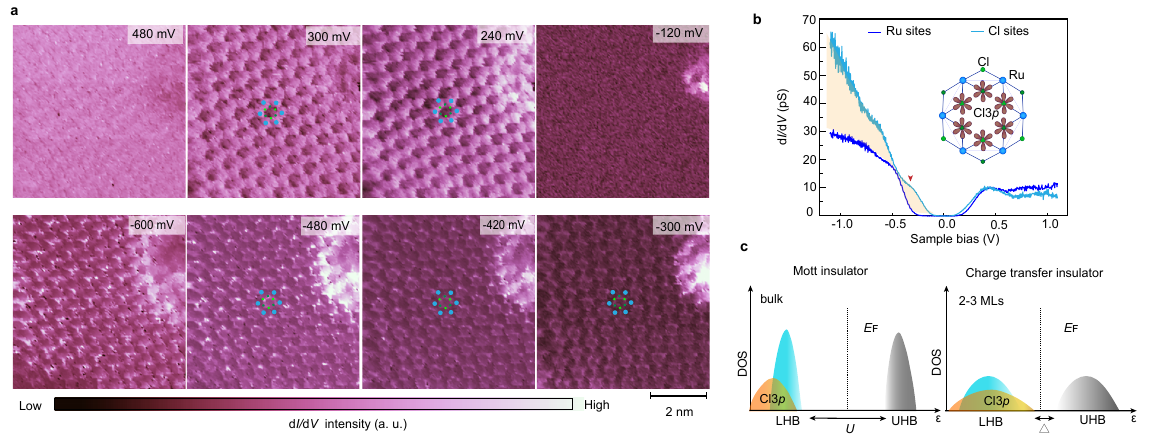}\\
 	\linespread{1}
 	\caption[Figure 3]{\textbf{Mott insulator (MI) to charge-transfer insulator (CTI) transition.} \textbf{a} Bias-resolved d$I$/d$V$ maps taken on the same region of 3-MLs $\alpha$-RuCl$_3$ ($V_{\rm bias}$=900 mV, $I_{\rm set}$=1 nA). Orbital textures corresponding to the Ru sites (light blue dots) are detected at the edge of UHB (180-300 mV). In contrast, the negative bias side displays the textures of Cl 3$p$ orbitals (green dots); \textbf{b} A comparison of averaged d$I$/d$V$ spectra taken on the Ru and Cl sites reveals enhanced occupied states and distinct dip-humps at Cl sites ($V_{\rm bias}$=800 mV,  $I_{\rm set}$=800 pA). Inset schematically shows the configuration and pattern of Cl 3$p$ orbitals; \textbf{c} Schematic DOS (density of states) from MI to CTI when the Cl 3$p$ orbitals enter the Mott gap of Ru 4$d$ electrons with changing thickness (layer-number). U(L)HB is an abbreviation of upper (lower) Hubbard band; $U$ is the Coulomb interaction and $\triangle$ denotes the charge transfer gap.}
 	\label{fig:figure-3}
 \end{figure}

We found that the super-modulation on lattice is absent for samples of 4-MLs and thicker. We took d$I$/d$V$ map for a thicker (Supplementary Fig. 1) and a 4-MLs samples(Supplementary Fig. 5), respectively. No super-modulations were observed in the surface DOS and the data are indistinguishable with the orbital textures. Subsequently, the surface morphology and LDOS were measured on a lower terrace (3-MLs), wherein the RG phase appeared but without any super-modulation. The bias-dependent d$I$/d$V$ maps are illustrated in Fig.\ref{fig:figure-3}a. The images exhibit orbital textures linked to the Ru $t_{2g}$ bands at the biases surpassing the UHB (240 and 300 mV). Within the sample biases of -300 to -480 mV, the textures in occupied states display a ring-shape pattern sharing the symmetry of the Cl arrangements (represented by green dots). At bias below -600 mV, the orbital textures converge towards the centre of the ring and become blurred. The ring shape pattern can be attributed to the 3$p$ orbitals, as shown in the inset of Fig.\ref{fig:figure-3}b. Quantitative analysis of the spectra obtained from the Ru and Cl sites reveals that the ring-shape texture of Cl 3$p$ orbitals corresponds to the shoulder peak in the d$I$/d$V$ spectra, as depicted in Fig.\ref{fig:figure-3}b.\par

The experimental detection of the Cl 3$p$ orbitals is crucial for understanding the reduction of Mott gap. First, notice that part of the electrons in graphite are transferred towards the $\alpha$-RuCl$_3$ \cite{biswas_electronic_2019,mashhadi_spin-split_2019,zhou_evidence_2019,rizzo_charge-transfer_2020,rizzo_nanometer-scale_2022}. If these electrons are doped into the Mott-Hubbard bands of $\alpha$-RuCl$_3$, they will induce a finite DOS at zero bias (FL) by causing double occupancies and charge carriers (i.e. the doublons). Nevertheless, a (reduced but finite) charge gap was persistently observed, which leads to the conclusion that the electrons do not enter the Mott-Hubbard bands; rather, they formed an interfacial dipole layer as a result of the difference in chemical potential between $\alpha$-RuCl$_3$ and graphite \cite{rossi_direct_2023}. The dipole layer will significantly change the energy levels of the ions, especially the outer shell 3$p$ orbitals of anions which are less screened (in contrast, the energy levels of the 4$d$ orbitals of the cations are less affected due to screening of the outer shell orbitals) \cite{souza_magnetic_2022,biswas_electronic_2019}. Apparently, for 2-MLs and 3-MLs, the energy levels of the Cl 3$p$ orbitals are pushed into the Mott gap between 4$d^5$ and 4$d^6$ of the Ru$^{3+}$, which turns the flake of $\alpha$-RuCl$_3$ into a CTI with dramatically reduced gap, as schematically shown in Fig.\ref{fig:figure-3}c. The transition from MI to CTI is depicted in the form of additional low energy states on Cl sites (Fig.\ref{fig:figure-3}b), as well as the noticeable shift in band-edges toward FL as the surface approached the heterointerface (Fig.\ref{fig:figure-1}d). On the other hand, in contrast to the situation depicted in Fig.\ref{fig:figure-3}, we did not observe any orbital textures originating from Cl 3$p$ in the d$I$/d$V$ maps of 1-ML case, even though there are in-gap states in the d$I$/d$V$ spectrum \cite{zheng_insulator--metal_2024}. Here, we 
propose that the enhanced crystal field due to the possible buckling of Ru atoms in 1-ML \cite{yang_magnetic_2023} and the strong electric field by the charge transferred from graphite to the surface \cite{biswas_electronic_2019}
restructure the orbital configurations in $\alpha$-RuCl$_3$ \cite{wang_identification_2022}, keeping 1-ML a MI. For 4-MLs and above, the electric field reduces quickly such that the energy level of Cl 3$p$ orbital is not close to the FL, hence the charge gap is not significantly reduced. \par

After the finish of the current work, a preprint \cite{kohsaka_imaging_2024} reported the observation of quantum oscillations at 4.2 K on 1-ML $\alpha$-RuCl$_3$ that was directly grown on graphite. The sample in \cite{kohsaka_imaging_2024} exhibits a reduced Mott-gap of around 0.6 eV, 
and the STM images also reveal the super-modulation patterns extending beyond the oscillation around the defects, as  discovered in our 2-ML samples. Noticing that the growth 1-ML $\alpha$-RuCl$_3$ on graphite is thicker than the transferred 1-ML sample, this may be the reason that super-modulation was not observed in our 1-ML sample. \par

\section*{Discussion}
The charge super-modulations in electron Fermi liquids, which have a large electron FS, are often linked to CDW, moir\'{e} patterns, or quasiparticle interference (QPI). In the case of the few-layer $\alpha$-RuCl$_3$ on graphite where a full charge-gap is present, these mechanisms can all be ruled out. The Peierls-type CDW and correlation-driven charge order, if it takes place, would appear at considerably lower energy \cite{frano_charge_2020}. In addition, the formation of moir\'{e} patterns is implausible due to the significant lattice mismatch between $\alpha$-RuCl$_3$ and graphite \cite{biswas_electronic_2019}, as is already confirmed by the absence of the super-modulation in 1-ML case in our recent experiments \cite{zheng_tunneling_2023, zheng_insulator--metal_2024}. The observation of comparable super-modulation on 1-ML $\alpha$-RuCl$_3$ growth directly on graphite in ref. \cite{kohsaka_imaging_2024} can further effectively exclude the moir\'{e} patterns formed due to a twist angle between two $\alpha$-RuCl$_3$ layers. Furthermore, due to the lacking of bias dependence, the observed super-modulation is inconsistent with the QPI patterns as well \cite{avraham_quasiparticle_nodate}. Finally, with charge transferred from graphite to heterointerface, the nesting between two hole-pockets at the positions of Dirac cones ($K$ and $K^{'}$) with wave vector $\mathbf k$ may affect the charge distribution in layered $\alpha$-RuCl$_3$. However, this can$^{'}$t interpret the fact that wave vector of the charge super-modulation in 2-MLs $\alpha$-RuCl$_3$ is independent on the relative orientation of the substrate. \par

  \begin{figure}[htp]
	\centering
	\includegraphics[width=1\columnwidth]{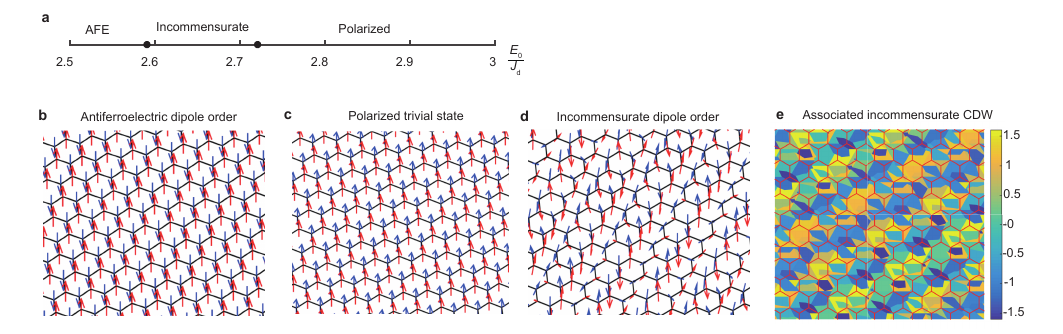}\\
	\linespread{1}
	\caption[Figure 4]{\textbf{Anti-ferroelectricity and the incommensurate dipole order.} \textbf{a} Phase diagram of the dipole model, where `AFE' stands for the anti-ferroelectric phase. Cartoon picture of the dipole distribution in three phases: \textbf{b} The anti-ferroelectric phase (the blue/red arrows stand for A/B-sublattice respectively); \textbf{c} The polarized trivial phase; \textbf{d} The incommensurate dipole order phase. \textbf{e} Illustration of the charge density wave(CDW) order associated with the incommensurate dipole order.}
	\label{fig:figure-4}
\end{figure}

A plausible interpretation of the super-modulation is from the interference of Majorana spinons. At temperatures slightly above the N\'{e}el temperature $T_c$, the material $\alpha$-RuCl$_3$ in 2D limit is proximate to a Kitaev spin liquid owing to the combined thermal and quantum fluctuations. Especially, in a temperature region the thermally excited $Z_2$ fluxes yield a finite Fermi surface of the Majorana spinons, which are called thermal Majorana metal in literature \cite{yoshitake_fractional_2016,yoshitake_majorana_2017, eschmann_thermodynamics_2019,nasu_nonequilibrium_2019,nasu_thermodynamic_2020, lee_kondo_2023}. The nesting of the Majorana Fermi surface can generate a spinon density waves. The internal coupling between spinons and chargons (i.e. holons or doublons depending on the bias voltage) will potentially cause a charge order and quantum oscillation of the electron density \cite{chen_evidence_2022}. However, the size of Majorana fermi-surface should be temperature dependent, which is inconsistent with the fact that the super-modulations are detectable at both temperatures of 77 K and 4.2 K with the same wave vector \cite{kohsaka_imaging_2024,qiu_evidence_2024}. At this point, there exist no clear evidences to establish a connection between the observed super-modulation and Majorana fermions.\par 

Here we propose an alternative mechanism that is closely related to the anti-ferroelectricity of $\alpha$-RuCl$_3$ caused by the charge hopping between anions and cations in the CTI in an inhomogeneous manner \cite{khomskii_transition_2014, Pavarini_819465}. At zero electric field, the electric dipole moments are oriented anti-ferroelectrically along the $c$-direction. However, the pattern of dipole order can be changed by the strong internal electric fields from the charges transferred to the interface or at vacancy defects. To investigate the effect of electric field, we consider an effective model of the electric dipoles (Supplementary note 4) containing dipole-dipole interaction $H_{1}$ with strength $J_d$, anisotropic interaction $H_{2} = \sum_{\langle i,j\rangle\in (\alpha\beta)\gamma} [K p_i^\gamma p_j^\gamma + \Gamma (p_i^\alpha p_j^\beta + p_i^\beta p_j^\alpha)]$ resulting from charge-spin coupling, easy-axial anisotropy and internal electric field $H_{3} = \sum_i[D(\mathbf p_i\cdot \hat c)^2 - E_i \mathbf p_i\cdot \hat c]$, assuming that the Fourier component of the electric field obey Gaussian distribution in momentum space $E_k = E_0 e^{-{k^2\over \alpha^2}} $ (with $\alpha=0.02$ plus an anisotropy). Adopting the parameters $ D/J_d=-1.28, K/J_d=0.1, \Gamma/J_d = -0.4$, we obtain the phase diagram shown in Fig.\ref{fig:figure-4}a. Due to the easy-axial anisotropy potential $D$ and the dipole-dipole interactions $J_d$, the dipoles favor an anti-ferroelectric ground state (Fig.\ref{fig:figure-4}b) when the electric field $E_0$ is weak $E_0/J_d < 2.59$. When the electric field is strong $E_0/J_d>2.72$, the ground state will be turned into a polarized state (Fig.\ref{fig:figure-4}c). However, when the electric field falls in an intermediate window $2.59<E_0/J_d<2.72$, the dipoles form incommensurate order with the wave vector \textbf{k}$\sim 0.40$ r.l.u. (see Fig.\ref{fig:figure-4}d) which is very close to \textbf{k}$_{\rm sm} $=0.39 r.l.u. of the experimentally observed super-modulations. The incommensurate dipole order $\mathbf p(r)$ induces a CDW order $\rho(r)$=$\nabla\cdot\mathbf p(r)$ with the same wave vector $\mathbf k$ (as shown in Fig.\ref{fig:figure-4}e). This provides a possible origin of the super-modulations in the STM experiment. In the experiment, the electric field $E_0$ is generated by the interfacial dipole between $\alpha$-RuCl$_3$ and graphite, and it decays on $\alpha$-RuCl$_3$ surface as the thickness increases. Consequently, the intralayer dipoles of $\alpha$-RuCl$_3$ form the incommensurate order when the electric field falls in the certain window, as evidenced by 2-ML sample. We infer that $E_0$ on 3-ML case has been outside that range with $E_0/J_d < 2.59$, and the intralayer dipoles favor an anti-ferroelectric ground state. The incommensurate dipole order may also contain the contribution of lattice distortion due to softening of certain phonon mode caused by electric force between the substrate and the $\alpha$-RuCl$_3$ layer \cite{li_giant_2021, metavitsiadis_optical_2022}. \par

In summary, we performed systematic STM/STS study of few-layer $\alpha$-RuCl$_3$ proximate to a graphite surface. Our data indicate that $\alpha$-RuCl$_3$ is changed from a bulk MI to a CTI (2-MLs and 3-MLs) due to the shift of energy levels of the 3$p$ orbitals in the Cl$^-$ anions, which guarantees the charge excitation near the Fermi level. Furthermore, exotic incommensurate super-modulations of charge states were observed in 2-MLs samples, which is here consistently interpreted as a charge order associated with an incommensurate electric dipole order. The findings highlight the significance of anions in tuning the Mott-Hubbard bands in Kitaev materials, and indicate that the strong coupling between charge and spin degrees of freedom, which may result in electrical accessible method of detecting the magnetic properties. Our work urge the experimental studies to further explore the electric dipole degrees of freedom in $\alpha$-RuCl$_3$ and to discover the microscopic mechanism of the interactions between charge and spin. \par

\section*{Methods}
In this work, we exfoliated the $\alpha$-RuCl$_3$ thin flakes using the scotch tapes, from a bulk crystal synthesized by vacuum sublimation of commercial high purity (99.99\%) $\alpha$-RuCl$_3$ powder. After repeatedly reducing the thickness of the film, we exfoliated again using a thermal release tap. Then the thermal release tap with the $\alpha$-RuCl$_3$ thin flakes was attached on a fresh surface of HOPG substrate. After heating the sample to 120 $^o$C for 20 seconds in the atmosphere, the $\alpha$-RuCl$_3$ thin films were released on the HOPG surface. Then the sample was annealed at 280 $^o$C in ultra-high vacuum chamber (1E-10 Torr) in STM system for at least 2 hours for degassing and improving the contacting quality, before STM/STS measurements. We measured $\alpha$-RuCl$_3$ flakes with different thicknesses randomly, and captured the STM topographic images and the STS spectra on the chosen regions. In this experiment, commercial STM system (Unisoku-1300) has been used at a base temperature of 77 K. The STM tip is made by tungsten wire with self-corrosion in 4 mol/L NaOH. The STS spectra measurements were carried out using a lock-in technique at a frequency of 707 Hz and a modulation voltage ranges from 5 mV to 10 mV.\par

\section*{Data Availability}
All data supporting the findings of this work are presented in the manuscript and its associated Supplementary Information. Additional data are available from the corresponding author(s) upon request. Source data are provided with this paper.

\section*{Acknowledgments}
X.H.Z. is supported by funding from Beijing Academy of Quantum Information Sciences; R.R.D. is support by the funding from National Basic Research and Development plan of China (Grants No. 2019YFA0308400) and Innovation Program for Quantum Science and Technology (Grant No, 2021ZD0302600); Z.X.L. thank the support from National Basic Research and Development plan of China (Grants No.2023YFA1406500, 2022YFA1405300) and NSFC (Grants No.12374166, 12134020); K.T. is supported by the National Natural Science Foundation of China (Grants No.12174027); Y.G.S. is supported by the National Natural Science Foundation of China (Grants No. U2032204, U22A6005) and the Synergetic Extreme Condition User Facility (SECUF).
\section*{Author contributions}
X.H.Z., Z.X.L. and R.R.D. supervised the project. C.W.Z., H.X.Z., C.L.Y. and Y.G.S. synthesized the single crystal. X.H.Z. prepared the layered $\alpha$-RuCl$_3$ on graphite and performed STM/STS measurements. Z.X.L. developed the theoretical model and performed the simulation. X.H.Z., Z.X.L., K.T. and R.R.D. analyzed and visualized the data. X.H.Z., Z.X.L. and R.R.D. wrote the manuscript with input from all authors. All authors discussed the results and contributed to the manuscript. 
\section*{Competing interests}
The authors declare no competing interests.
\section*{Additional information}
\end{document}